# How a blister heals

Jonathan E. Longley[1], L. Mahadevan[2] and Manoj K. Chaudhury[1*]

[1]Department of Chemical Engineering, Lehigh University, Bethlehem, PA 18015
[2]School of Engineering & Applied Sciences, and Department of Physics, Harvard University,
Cambridge, MA
02138, USA

## Abstract

We use experiments to study the dynamics of the healing of a blister, a localized bump in a thin elastic layer that is adhered to a soft substrate everywhere except at the bump. We create a blister by gently placing a glass cover slip on a PDMS substrate. The pressure jump across the elastic layer drives fluid flow through micro-channels that form at the interface between the layer and the substrate; these channels coalesce at discrete locations as the blister heals and eventually disappear at a lower critical radius. The spacing of the channel follows a simple scaling law that can be theoretically justified, and the kinetics of healing is rate limited by fluid flow, but with a non-trivial dependence on the substrate thickness that likely arises due to channelization. Our study is relevant to a variety of soft adhesion scenarios.

*mkc4@lehigh.edu

**Introduction.**

Blisters, blebs and boils are protrusions of a skin that separates from a solid substrate. They are ubiquitous in science and technology, and arise in situations ranging from the pesky bubble that is hard to a eliminate from wall-paper to the localized delaminations in thin adhesive layers[1], from the active membrane protrusions in a cell [2] to the fluid-filled blisters in inflamed tissues and their physical counterparts in geophysics. How they nucleate, grow and eventually saturate is now a well studied problem; following nucleation at a defect or debonded region, the growth of a blister is determined by the balance of the internal pressure, adhesive forces, mode mixity, film elasticity, and the deformation of the film and/or the substrate [3-10]. Indeed, the mechanism by which blisters form due to internal pressure is now a standard method to quantify the work of adhesion between the film and the substrate, a fundamental property of the interface. Once a blister forms, it does not always maintain itself and can slowly heal as the fluid permeates away from the localized zone of high pressure. However, how a blister heals does not seem to have been studied previously. Here, we investigate this process using a simple experiment that follows the disappearance of a blister trapped between a thin glass cover slide and a PDMS film, relevant for any system where the escape or removal of fluid is essential to form a uniform adhesive bond.

**Experimental Methods**

Our minimal system consists of an elastic plate (glass cover slide) that is gently placed onto a soft sticky substrate (10-240 μm thick PDMS film). As the crack between the cover slide and the PDMS film heals (which occurs spontaneously after contact), the edge of a razor blade is used to guide the crack front to trap a pocket of air and form a blister that is initially axisymmetric. The evolution of the size and shape of the blister is recorded using an optical microscope equipped with a monochromatic light source (Figure 1a), using interference fringes (Figure 1b) to estimate the height profile of the blister, knowing that constructive interference bands occur at heights of $\frac{(2n-1)\lambda}{4}$ and destructive interference at $\frac{n\lambda}{2}$, where n=1,2,3..... and λ is the wavelength of light.

We find that fingering instabilities form on the blister periphery for all film thicknesses. However, as shown in Figure 1b, once the blister shrinks to a critical size, the distal tips of the fingers become pinned even as the fingers grow inward and the central part of the blister shrinks rapidly. This process forms a set of interconnected channels which display multiple branches. To understand the origin of this branching, we consider the effects of blister radius and PDMS film thickness on the fingering wavelength, as shown in Figure 1c, and find that the wavelength of the fingering instability is independent of radius and increases linearly with substrate thickness. This is consistent with previous experimental and theoretical studies [11-18] of fingering instabilities in a confined soft elastic layer that find the characteristic wavelength $\lambda = 3.7H$.

Further examination of the finger wavelength in the final snapshot of Figure 1b reveals that the wavelength changes discretely with radius and the formation of fingers occurs as the system attempts to maintain the same finger wavelength at different radii from the blister center, with the ends of the finger remaining pinned. By observing the change in the outer radius of the blister (measured at the ends of the fingers) over time vs. the inner radius (measured at the base of the fingers) this pinning effect becomes apparent as shown in Figure 2a. The outer radius of the blister (blue circles, Figure 2a) displays a stick-slip motion; the fingers are sequentially pinned, stretched, and broken due to the shrinkage of the blister. Using interference patterns, we show the height profile of the blister in Fig. 2b, using a simple model for the deflection of the glass cover slip as discussed in the next section.

In the branching pattern formed just prior to blister disappearance the channel width varies as a function of radius and decreases with increasing radius as shown in Figure 2c. This is expected if the air escapes through channels formed at the glass/PDMS interface to accommodate the flux as a function of radius. The relevance of this observation will become clear in the next section when we consider the mechanism the air utilizes to escape from the blister.

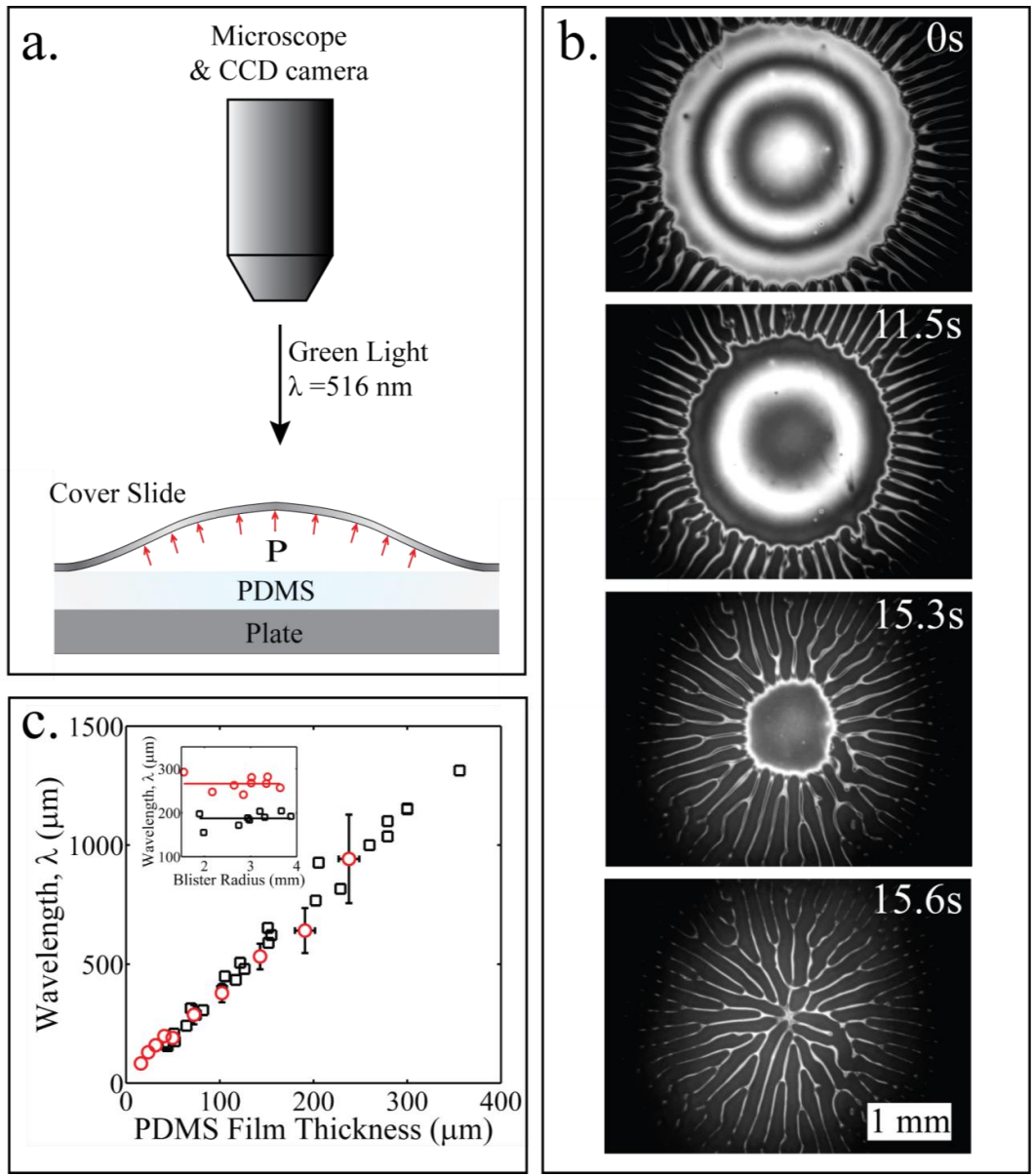

**Figure 1:** (a) Schematic of experimental set up. (b) Micrographs taken during the final stages of the life of a blister trapped between a glass cover slip and 50 μm PDMS film. The formation of a multi-branched network is observed in the final stage. (c) Wavelength of the fingering instability as a function of PDMS film thickness. The length used to calculate the finger wavelength is measured along the ends of the fingers. Red circles represent the finger wavelengths calculated for the blister experiments. For the thicknesses measured below 100μm error bars are less than the size of the marker. Black squares represent the finger wavelength data collected by Ghatak et al[11] for two different geometries. Insert: Fingering wavelength as a function of blister radius for PDMS films of thicknesses 75 μm (red circles) and 50 μm (black squares). Fingering wavelength remains constant as the blister shrinks.

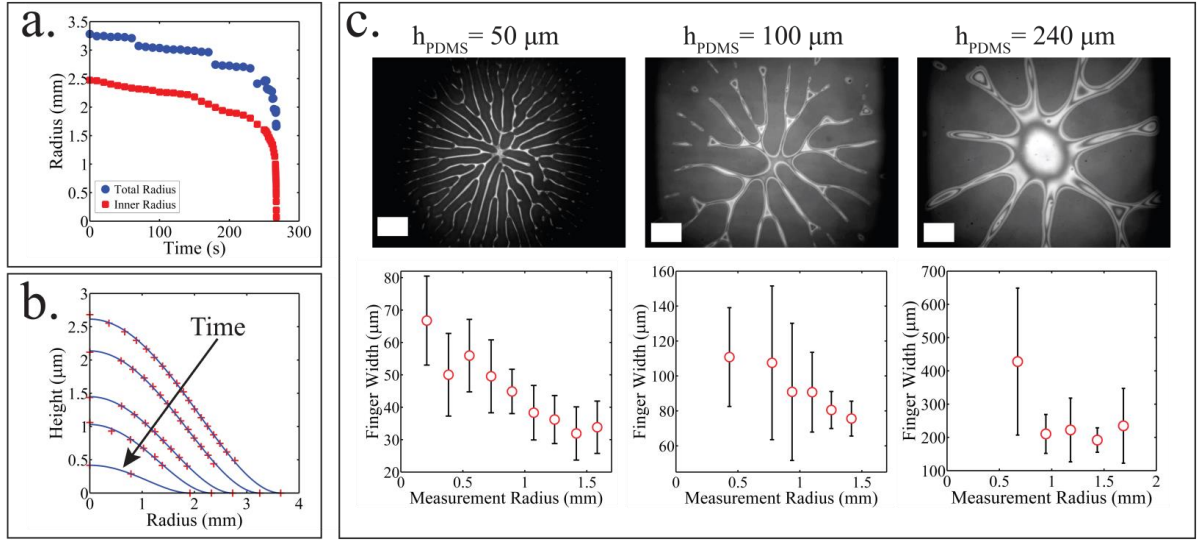

**Figure 2:** Quantification of the blisters and fingers produced in the final stages of a blister's lifetime. (a) Outer radius of the blister (blue circles, measured at the ends of the fingers) and inner radius (red squares, measured at the base of the fingers) as a function of time for the healing of a blister on a 50 μm thick PDMS film. In the final stages of blister life the outer radius remains constant while the inner radius rapidly decreases. (b) Height profiles of a blister on a 50 μm thick PDMS film plotted at different instances in time. The height profile, red crosses, is determined from the interference fringes observed when viewing the blister under monochromatic light (λ=516 nm). A simple model (blue lines) which only considers the deformation of the glass cover slide is used to fit the data. (c) Micrograph snapshots taken just before blister disappearance for experiments on 50, 100 and 240 μm thick films. White scale bars represent 0.5 mm. The corresponding finger width as a function of measurement radius is shown beneath each micrograph. For 50 and 100 μm thick PDMS films the finger width decreases as we move away from the blister center.

**Results and Discussion.**

**Analysis of the blister height profile and estimation of its internal pressure**

To understand our experimental observations, we note that the dynamics of healing is determined by a balance between plate deformation and fluid flow in the gap. Since the time scale over which the process occurs is large relative to the time scale for inertial waves in the plate, and inertial forces in the fluid are dominated by viscous forces, we consider the quasi-static limit for elastohydrodynamics wherein the cover slip may be approximated by a thin plate undergoing a small bending deformations (vertical deformation ~ plate thickness) with its shape determined by the equation [19],

$$B\nabla^4 \zeta = -P \qquad (1)$$

where $B$, $\zeta$, $P$ are the plate flexural rigidity, vertical plate displacement and pressure respectively. Solving Equation (1) assuming axisymmetric deformations (ignoring the small variations due to the presence of channels) with the boundary conditions $\zeta = 0$ and $\frac{d\zeta}{dr} = 0$ at r=R, the boundary of the blister leads to the height profile,

$$\zeta = \frac{P}{64B}(R^2 - r^2)^2 \qquad (2)$$

Figure 2b shows the model fit (blue lines) for height profiles obtained at different times for a blister on a 50 μm thick film, with P, R treated as adjustable parameters. Close inspection of the fingers show that the interference fringes begin at the end of the finger. Due to the sharp deformation of the PDMS at the end of the finger it is not possible to directly measure the number of fringes in the finger which correspond to the plate deflection. Therefore, in the analysis of a blister on a 50 μm thick PDMS film an approximate number of 5 interference fringes along the length of the finger is used.

The healing process is driven by the work of adhesion between the glass plate and the PDMS. As a first approximation to calculate the energy release rate of the system we used the standard result of Williams for a circular blister [4], $G = \frac{\Delta P^2 R^4}{128B}$, which can also be written in the same general form as Ombreioff's classical result for the peeling of mica [20], i.e.

$G = 32\frac{BH_{max}^2}{R^4}$. Here $G$, $\Delta P$, $R$ are the energy release rate, the pressure difference across the glass cover slide and the blister radius. This model only considers the bending deformation in the glass plate, i.e. any deformation in the PDMS is ignored, with the values for $\Delta P$ and $R$ obtained from the fitting Equation 2 to the blister profile. For very thin films ($h_{PDMS}$ =20μm) this model predicts the energy release rate to be, $G \sim 40$ mJ/m$^2$, which is close to the work of adhesion for a crack healing between a glass plate and PDMS film [17], $W \sim 44$ mJ/m$^2$. However, with an increase of the film thickness Williams's equation yields $G < W$; this discrepancy is due to the deformation in the PDMS at the crack tip that has not been accounted for in the simple model. Furthermore, although $G$ is weakly dependent on film thickness, for the analysis presented below, what is most important is that it is more or less independent of time.

**Modeling the escape of air from inside the blister to its surroundings**

One possible method for the air to escape is via diffuse flow through the PDMS film. The air has two possible directions for movement in the PDMS; it can diffuse into the film directly below the blister and/or diffuse through the annulus formed at the edge of the blister. If diffusion through the annulus is the rate controlling step, since the area of this annulus, $A = 2\pi R h_{PDMS}$, is linearly proportional to the film thickness, we expect that the permeability of the film should be proportional to the thickness of the substrate. Moreover, since the dynamics of the process are dominated by slow escape of air through the channels and the film, we model this via a rudimentary form of the Darcy's law for flow through an effective porous medium corresponding to the channels and the PDMS film via the relation,

$$\frac{dn}{dt} = -k_D \Delta P \qquad (3)$$

Where $\frac{dn}{dt}$ is the molar flow rate, $k_D$ is the permeability coefficient, $\Delta P = P_1 - P_2$, where $P_1$ and $P_2$ are the internal and the external pressures of the blister. Using (2), the volume of the blister can be expressed as $V = \int 2\pi r \zeta dr = \frac{\pi \Delta P R^6}{192 B}$. Substituting this relation for $V$ in light of the ideal gas law ($\underline{n} = P_1 V / \overline{R} T$) and Williams's equation ($G = \frac{\Delta P^2 R^4}{128 B}$) into (3) and rewriting in terms of $R$ yields,

$$[AR^5 + CR^3]\frac{dR}{dt} = -k_D \qquad (4)$$

Where $A = \frac{\pi P_2}{48 \overline{R} T B}$ and $C = \frac{\pi}{\overline{R} T}\sqrt{\frac{G}{72 B}}$, with $\overline{R}$ and $T$ being the molar gas constant and temperature. Equation 4 can be integrated to yield

$$\frac{A}{6}R^6 + \frac{C}{4}R^4 = -k_D t + D \qquad (5)$$

where $D$ is the constant of integration. In Figure 3, we plot $f(R) = \frac{A}{6}R^6 + \frac{C}{4}R^4 - D$ as a function of time to determine the permeability coefficient $k_D$ [21,22] which is seen to exhibit a strong dependence on the film thickness, Figure 3b, via a power law of the form $k_D \sim h_{PDMS}^{1.5}$. This is inconsistent with the simple analysis of diffusive flow through PDMS, but points towards a possible role of microchannels along the interface shown in Figure 2c in allowing air to flow through.

To permit the formation of microchannels the pressure inside the blister should be larger than a critical blistering pressure for the escaping air to lift the coverslip off the PDMS film. We estimate the average gauge pressure inside a blister to increase by about 0.1-0.3 Bar as the blister shrinks . If the thin PDMS film adhering to glass is not in perfect contact, it is possible that either the air escapes through this region or that this initial separation allows the air to create channels along the interface.

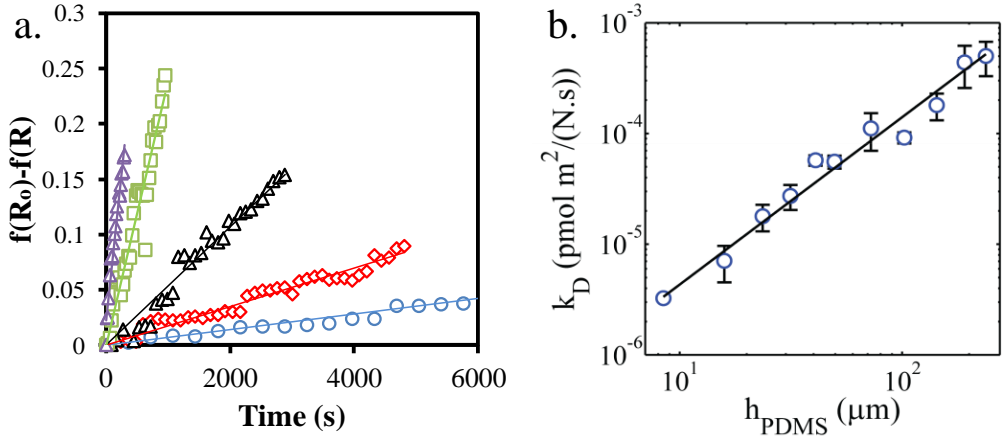

**Figure 3:** (a) Plot of f(Ro)-f(R) vs. time for a blister healing on PDMS films of different thickness. The function $f_1(R)$ is defined as the left hand side of equation 5. $f(R_0)$ is the value of the function at t=0. The blue circles, red diamonds, black triangles, green squares and magenta inverted triangles represent 15, 30, 40, 140, 240 µm thick PDMS films. The gradient of each least square fit is equal to the permeability coefficient, $k_D$, for that particular thickness (Equation 5). (b) Permeability coefficient, $k_D$, as a function of film thickness. The permeability coefficient increases with the film thickness obeying a power law trend of the form $k_D = p(h_{PDMS})^q$, where $p = 1.3 \times 10^{-7}$ (pmol m$^2$/(N.S))(um)$^{-q}$ and $q = 1.5$.

Our experiments and analysis are the first steps in understanding the innocuous question of how a blister heals – they show that although adhesion drives healing, the rate limiting step is the squeezing of fluid that occurs via micro-channels that form a branched structure. Our study captures the scaling dependence of the form of the blister, the channel spacing and the effective kinetics of healing.

An important question that remains unanswered is the dynamics of channelization as the blister heals – our experiments are unable to resolve the rapid kinetics of this process, and further work is required to address this question. Additionally, a fully coupled theory for the dynamics of the healing adhesion front that complements our parametric approach is a natural next question that needs to be addressed. In this context, it would also be interesting to study how viscous liquids are squeezed out by these types of self-generated generated flow channels. As the high viscosity liquids would exert a higher lubricating pressure, it should produce channels of larger hydraulic radius than those formed by the lower viscosity liquids. There could thus be an interesting compensation between high viscosity and high hydraulic radius so that the net squeeze flow rate is independent of viscosity. This may have some resemblance to the anomalous viscosity dependent lubricated friction in soft elastic contacts of the type reported recently [23].

**References**

1. Chaudhury M. and Pocius AV (eds) Surfaces, chemistry and applications: adhesion science and engineering (2002), Elsevier Science B.V, The Netherlands

2. Charras, G.T., Coughlin, M., Mitchison, T.J. and Mahadevan, L.; *Biophys. J.*, **94** (2008), 1836.

3. Dannenberg, H., *J. Appl. Pol. Sci.,* **5** (1961) 125.

4. Williams, M. L., *J. Appl. Pol. Sci.,* **13** (1969) 29.

5. Gent, A. N. and Lewandowski, L. H., *J. Appl. Pol. Sci.,* **33** (1987) 1567.

6. Briscoe, B. and Panesar, S., *Proc. Roy. Soc. London Ser. A.,* **433** (1991) 23.

7. Condon, J. B. and Schober, T., *J. Nucl. Mat.,* **207** (2007) 1.

8. Hutchinson, J. W. and Suo, Z. *Adv. Appl. Mech.,* **29** (1992) 63.

9. Hutchinson, J. W., Thouless, M. D. and Linger, E. G., *Acta Metall. Mater.,* **40** (1992) 295.

10. Zong Z., Chen, C-L., Dokmeci, M. R and Wan, K-T., *J. Appl. Phys.,* **107** (2010) 026104.

11. Ghatak, A. and Chaudhury, M., *Langmuir,* **19** (2003) 2621.

12. Ghatak,A. Chaudhury, M., Shenoy, V. & Sharma, A. *Phys. Rev. Lett,* **85**, 4329–32 (2000).

13. Shull, K., C. M. Flanigan, and A. J. Crosby, *Phys. Rev. Lett.,* **84**, 3057 (2000).

14. Ghatak, A., *Phys. Rev. E.*, **73** (2006) 041601.

15. Ghatak, A and Chaudhury, M. K., *J. Adhes*,. **83** (2007), 679.

16. Adda-Bedia, M. and Mahadevan, L. *Proc. Roy Soc. London. Series A.,* **462** (2006) 3233.

17, Ghatak, A., Mahadevan, L. and Chaudhury, M.K., *Langmuir,* **19** (2005) 1277.

18. Ghatak, A., Mahadevan, L., Chung, J. Y., Chaudhury, M. K. and Shenoy, V., *Proc. Roy Soc. London. Series A.,* **460** (2004) 2725.

19. Landau, L. D. & Lifshitz, E. M. *Theory of Elasticity*. 38–42 (Elsevier: 1986).

20. Obreimoff, J. *Proc. Roy Soc. London. Series A.,* **127** (1930) 290.

21. It is tempting to compare the permeability coefficients obtained here with the typical values reported in literature. This is, however, not a easy task as there is a large variation of the permeability values reported in literature. Note also that the relationship between the Darcy's permeability and the thickness of PDMS as reported in fig. 3b is empirical, which is dimensionally incomplete. A very crude comparison can, however, be done using a value of the permeability (*P*=179 Barrer) of nitrogen through ℓ~1 mm thick PDMS membrane [22]. This yields a comparable Darcy's permeability (P.ℓ) in our setting as $k_D$ ~ $10^{-4}$ pmol.m$^2$/Ns, which is in the range of the values obtained for the thicker films.

22. Tremblay, P., Savard, M. M., Vermette, J., Paquin, R., *J.Membrane Sci.*, 282 (2006) 245.

23. Lorenz, B.,Kric,B.A.,Rodriguez,N., Sawyer, W. G., Mangiagalli, P, and Persson, B.N. J.

    J. Phys.: Condens. Matter 25 (2013) 445013